\documentclass[10pt,preprint]{emulateapj}

\begin{document}
\title{The ACS LCID project. X. The Star Formation History of \objectname[]{IC~1613}: 
Revisiting the Over-cooling Problem\altaffilmark{1}}

\author{
Evan D. Skillman\altaffilmark{2},
Sebastian L. Hidalgo\altaffilmark{3,4},
Daniel R. Weisz\altaffilmark{5,6,7},
Matteo Monelli\altaffilmark{3,4},
Carme Gallart\altaffilmark{3,4},
Antonio Aparicio\altaffilmark{4,3},
Edouard J. Bernard\altaffilmark{8},
Michael Boylan-Kolchin\altaffilmark{9}, 
Santi Cassisi\altaffilmark{10},
Andrew A. Cole\altaffilmark{11},
Andrew E. Dolphin\altaffilmark{12},
Henry C. Ferguson\altaffilmark{13},
Lucio Mayer\altaffilmark{14,15},
Julio F. Navarro\altaffilmark{16},
Peter B. Stetson\altaffilmark{17}, and
Eline Tolstoy\altaffilmark{18}}

\altaffiltext{1}{Based on observations made with the NASA/ESA Hubble Space Telescope, 
obtained at the Space Telescope Science Institute, which is operated by the 
Association of Universities for Research in Astronomy, Inc., under NASA contract NAS 
5-26555. These observations are associated with program \#10505}
\altaffiltext{2}{Minnesota Institute for Astrophysics, University of Minnesota, 
Minneapolis, MN, USA; skillman@astro.umn.edu}
\altaffiltext{3}{Instituto de Astrof\'\i sica de Canarias. V\'\i a L\'actea s/n.
E38200 - La Laguna, Tenerife, Canary Islands, Spain;
shidalgo@iac.es, monelli@iac.es, carme@iac.es, aparicio@iac.es}
\altaffiltext{4}{Department of Astrophysics, University of La Laguna. V\'\i a L\'actea s/n.
E38200 - La Laguna, Tenerife, Canary Islands, Spain}
\altaffiltext{5}{Astronomy Department, Box 351580, University of Washington, Seattle, WA, USA;
dweisz@uw.edu}
\altaffiltext{6}{Department of Astronomy, University of California at Santa Cruz,
1156 High Street, Santa Cruz, CA, 95064}
\altaffiltext{7}{Hubble Fellow}
\altaffiltext{8}{Institute for Astronomy, University of Edinburgh, Royal
Observatory, Blackford Hill, Edinburgh EH9 3HJ, UK; ejb@roe.ac.uk}
\altaffiltext{9}{Astronomy Department, University of Maryland, College Park, MD, USA;
mbk@astro.umd.edu}
\altaffiltext{10}{INAF-Osservatorio Astronomico di Collurania,
Teramo, Italy; cassisi@oa-teramo.inaf.it}
\altaffiltext{11}{School of Mathematics \& Physics, University of Tasmania,
Hobart, Tasmania, Australia; andrew.cole@utas.edu.au}
\altaffiltext{12}{Raytheon; 1151 E. Hermans Rd., Tucson, AZ 85706, USA;
adolphin@raytheon.com}
\altaffiltext{13}{Space Telescope Science Institute, 3700 San Martin Drive, Baltimore, MD 21218, USA; 
ferguson@stsci.edu}
\altaffiltext{14}{Institut f\"ur Theoretische Physik, University of Zurich,
Z\"urich, Switzerland; lucio@physik.unizh.ch}
\altaffiltext{15}{Department of Physics, Institut f\"ur Astronomie,
ETH Z\"urich, Z\"urich, Switzerland; lucio@phys.ethz.ch}
\altaffiltext{16} {Department of Physics and Astronomy, University of Victoria, BC V8P 5C2, Canada;
jfn@uvic.ca}
\altaffiltext{17}{Dominion Astrophysical Observatory, Herzberg Institute of
Astrophysics, National Research Council, 5071 West Saanich Road, Victoria,
British Columbia V9E 2E7, Canada; peter.stetson@nrc-cnrc.gc.ca}
\altaffiltext{18}{Kapteyn Astronomical Institute, University of Groningen,
    Groningen, Netherlands; etolstoy@astro.rug.nl}

\begin{abstract}

We present an analysis of the star formation history (SFH) of a field near
the half light radius in the Local Group 
dwarf irregular galaxy IC~1613 based on deep {\it Hubble Space Telescope} 
Advanced Camera for Surveys imaging. 
Our observations reach the oldest main sequence turn-off, allowing a time 
resolution at the oldest ages of $\sim 1$ Gyr. 
Our analysis shows that the SFH of the observed field in IC~1613 is 
consistent with being constant over the entire lifetime of the galaxy.
These observations rule out an early dominant episode of star formation
in IC~1613.  
We compare the SFH of IC~1613 with expectations from cosmological models.  
Since most of the mass is in place at early times for low mass halos, a naive expectation
is that most of the star formation should have taken place at early times.  Models
in which star formation follows mass accretion result in too many stars formed 
early and gas mass fractions which are too low today (the ``over-cooling problem'').  
The depth of the present photometry of IC~1613 shows that, at a resolution of $\sim 1$ Gyr,
the star formation rate is consistent with being constant, at even the earliest times, 
which is difficult to achieve in models where star formation follows mass assembly.  
\end{abstract}


\section{INTRODUCTION}\label{secint}

\subsection{Motivation}

The present paper is part of the Local Cosmology from Isolated Dwarfs 
(LCID)\footnote{Local Cosmology from Isolated Dwarfs: http://www.iac.es/project/LCID/} project. 
We have obtained deep {\it Hubble Space Telescope} (HST) photometry,
reaching the oldest main sequence turn-off points, of six isolated 
dwarf galaxies in the Local Group: IC~1613, Leo~A, 
Cetus, Tucana, \objectname[]{LGS-3}, and Phoenix. 
Five galaxies were observed with the Advanced Camera for Surveys \citep[ACS,][]{ford98}, 
while Phoenix was observed with the Wide Field and Planetary Camera-2 
\citep[WFPC2,][]{holtzman1995}. 
The main goal of the LCID project is to derive the 
star formation histories (SFHs), age-metallicity relations (AMRs), variable star 
populations, and stellar population 
gradients of this sample of galaxies. Our objective is to study their evolution 
at early epochs and to probe effects of cosmological processes, such as the cosmic 
UV background subsequent to the onset of star formation in the universe or 
physical processes such as the gas removal by supernovae (SNe) feedback. 
Our sample consists of field dwarfs which were chosen in an effort to study systems 
as free as possible from environmental effects due to strong interactions with 
a host, massive galaxy. 

The SFH is a powerful tool to derive fundamental 
properties of dwarf galaxies and their evolution \citep{tolstoy2009}, but to study 
the earliest epochs of star formation, deep CMDs, reaching the oldest main sequence 
turn-offs, are required \citep[cf.,][]{gallart2005}.  Our impressions of the evolution of 
dwarf galaxies are biased by studies of the nearby, gas-poor dSph companions 
of the Milky Way.  Because of their
proximity, deep CMDs are obtainable from ground-based observatories.  In contrast,
the gas-rich, dIrr galaxies, which are found at greater distances, have few
studies with comparable photometric depth.  The Magellanic Clouds represent
the one exception to this generalization, but, because of they are 
substantially more massive than the typical dwarf galaxy and 
have potentially complex histories due to their current interactions 
\citep[see][and references therein]{kalli2013}, they present less than ideal
targets for study.

Because of the larger distances to the dIrrs, until the LCID project, none have had 
resolved star studies which reach down to the oldest main sequence turnoff stars.
Thus, our view of the SFHs of dwarf irregular galaxies has been shaped by
relatively indirect measures.  The work by \citet{gallagher84} represents a seminal contribution to
our understanding of the SFHs of dIrrs.  Based on galaxy mass estimates, 
blue luminosities, and H$\alpha$ luminosities, they demonstrated that most irregular 
galaxies were consistent with nearly constant star formation over their lifetimes
(as opposed to the larger spiral galaxies which showed declining star formation rates, 
hereafter SFRs).  Most
importantly, they pointed out that ``The constant SFR history implies that the 
simple classical model in which star formation is proportional to gas density 
in a closed system cannot be correct for irregular galaxies.''  Modern observations
of nearby dIrrs have been collected and summarized by \citet{weisz2011},
and indicate that, on average, SFRs are higher at early times, but that there are 
some dIrrs for which constant SFRs are a good approximation.

Recently, there have been theoretical papers emphasizing the difficulty in
making models of galaxies which have nearly constant SFRs.
\citet{orban08}, \citet{sawala2011}, \citet{weinmann2012}, and \citet{kuhlen12} 
have all highlighted the difficulty of producing dwarf galaxies with properties 
comparable to those observed
in the current universe.  The degree of failure is greatest in the amount of
mass converted into stars, which is of order one magnitude too large in 
cosmological simulations.  This overproduction of stars in dwarfs is an 
extreme symptom of the ``over-cooling problem'' \citep[c.f.,][]{benson03} 
faced by all galaxy modeling.

Without deep imaging of resolved stars in dIrr galaxies, we lack sufficient
time resolution to study SFHs at the earliest times. For example, the earliest
time bin used by \citet{weisz2011} is 4 Gyr in duration.  At this 
time resolution,  we cannot distinguish
between SFRs that are constant at all times, or SFRs that show considerable 
variation (most importantly evidence for an early dominant episode of SF).
The initial time bin of 4 Gyr covers the time range from before re-ionization up
to a redshift of $\sim$2 \citep[which is approaching the peak of in the history 
of universal star formation, e.g.,][]{madau1998},
and represents this entire important range with a single average number.
The observations of IC~1613 presented here resolve this initial period and 
represent a small step into this relatively unexplored territory.

\subsection{The Normal, Isolated, Low Mass, Dwarf Irregular Galaxy IC 1613}

Here we present our analysis of a deep HST/ACS observation of a field in IC~1613.
As we discuss in section \ref{repfield}, the SFH that we derive for this
field is likely a good representation for the entire galaxy.
IC~1613 is a low-luminosity, Local Group, dIrr galaxy 
which is one of the nearest
gas-rich irregular galaxies \citep[for a review of the properties of
IC~1613 see][]{vdb2000}. 
Because of its proximity, IC~1613
offers the opportunity to reconstruct a detailed
SFH of a relatively isolated and non-interacting dwarf irregular galaxy.
IC~1613 also has very low foreground and internal reddening (although it lies
within 5 degrees of the ecliptic). There
have been several determinations of its distance based on Cepheid and RR~Lyr 
variable stars and the tip of the red giant branch. 
Using all three methods, \citet{dolphin2001} derived a distance of
730 kpc. In a comparison of all of the literature
values (which included their own measurement using RR Lyrae and Cepheid data
 of 770 Kpc), \citet{bernard2010} determine a mean
distance of 760 kpc. 
The most recent determination is by \citet{tammann2011} who derived 
a distance of 740 kpc using Cepheids.
For consistency with the other LCID studies, we will adopt the distance estimate
from \citet{bernard2010} which corresponds to scales of 221 pc arcminute$^{-1}$,
3.7 pc arcsecond$^{-1}$, and $\sim$ 0.2 pc pixel$^{-1}$.

IC~1613 is not considered to be a satellite of either the Milky Way or M31.
\citet{mateo1998} included IC~1613 in the   
diffuse ``Local Group Cloud,''  and \citet{mcc2012} determined a distance of
517 kpc and a velocity of $-$90 km s$^{-1}$ relative to the Local Group 
barycenter.  As such, it is located very close to the zero velocity surface
for M~31 and well within the zero velocity surface for the Local Group
\citep{mcc2012}.
Given its isolated position and velocity, IC~1613 has not had any recent 
interactions, although interactions with other galaxies long ago cannot be 
ruled out.
To date, there are no proper motion studies of IC~1613, which would be very
valuable in determining its potential interaction history.

The physical parameters of IC~1613 were summarized in \citet{cole1999}. 
These properties are normal for an Im~V galaxy with  a
moderate luminosity ($M_V = -$15.2) and a small value of the maximum
amplitude of the rotation curve \citep[V$_{max}$ $=$ 25 km s$^{-1}$;][]{lake1989}.
Its SFR of 0.003~$M_{\odot}$~yr$^{-1}$ \citep{mateo1998}
is also normal for its type and luminosity.
The most recent ISM oxygen abundance measurement was made by
\citet{lee2003}, who obtained a spectrum where the $\lambda$4363 auroral
line of [\ion{O}{3}] was detected resulting in a measurement of 
12 $+$ log (O/H) $=$ 7.62 $\pm$ 0.05.
This corresponds to 8.5\% of the solar oxygen abundance 
\citep[as determined by][]{asplund2009}, which
is slightly less than that of the SMC, and normal for a galaxy of its 
luminosity \citep[e.g.,][]{skillman1989, berg2012}.
This combination of proximity and normality makes IC~1613
one of the best opportunities to study the properties of a dwarf star-forming
galaxy that is relatively isolated (as is typical for Im~V galaxies).

The overall structure of IC~1613 is also typical of an Im~V galaxy.
Ground-based studies have provided an overview of the
stellar population distributions.
\citet{bor2000} studied the distribution of luminous cool stars from J and 
K-band imaging and found AGB stars covering a wide range in age in all
of their inner galaxy fields.  \citet{albert2000}
conducted a wide field survey of IC~1613 for C and M stars, and
found the (intermediate age) C stars extended out to 15 arcminutes, 
well beyond the regions where star formation currently is active. 
Recently, \citet{bernard2007} have conducted a wide field optical
survey of IC~1613, and trace red giant branch (RGB) stars out to radii
$\ge$ 16.5 arcminutes ($\sim$ 3.6 kpc), showing the galaxy to be more extended 
than previously thought.

The resolved stellar populations of 
IC~1613 have been studied with the HST twice in the past, both times using
the WFPC2 camera.
\citet{cole1999} studied a central field and found IC~1613 to be a
smoothly evolving galaxy with a relatively constant SFR over the last Gyr.
Horizontal branch (HB) stars were detected, indicating the presence of an old
population.
\citet{skillman2003a} obtained deep imaging for a field located 7.4 arcminutes
southwest of the center.  While that imaging was not quite deep enough to reach to the 
oldest main sequence turnoff stars, greatly limiting the time resolution at the
oldest ages, the derived SFRs were constant within a factor
of three over the entire lifetime of the galaxy.

In this paper, we present the SFH of IC~1613 obtained from observations with the ACS 
on the HST. The photometry reaches the oldest main sequence turn-offs of the galaxy, 
allowing us to obtain an accurate SFH even for the oldest stellar populations.
\citet{bernard2010} have already used these observations to conduct a study of 
the variable star content of IC~1613.

The structure of the paper is as follows: in \S\ref{secred} the observations and 
data reduction are discussed and the CMD is presented.
The derived SFH of IC~1613 is presented in \S\ref{secsfhresult} 
and is compared with those of other LCID galaxies in \S\ref{seclcid}. 
The implications of the SFH of IC~1613 for galaxy modeling, and, in particular,
the over-cooling problem 
are discussed in \S\ref{seccosmo}. The main conclusions of the work are summarized 
in \S\ref{seccon}.
As with the previous LCID papers, cosmological parameters of 
$H_0=70.5\rm~km~s^{-1}~Mpc^{-1}$, $\Omega_m=0.273$, and a
flat Universe with $\Omega_\Lambda = 1 - \Omega_m$ are assumed
\citep[i.e.,][]{kom_etal2009}.

\begin{figure}[h]
\epsscale{1.0}
\includegraphics[width=8.6cm,angle=0]{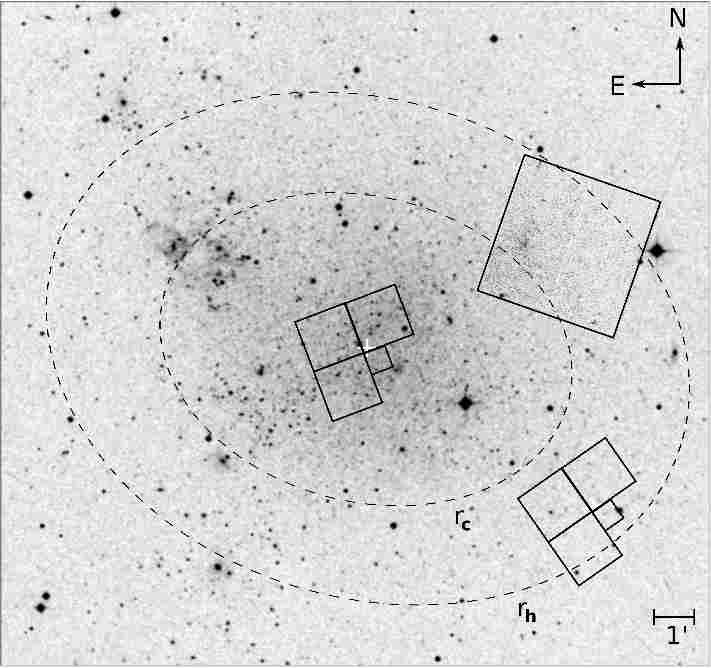}
\caption[ ]{The location of the newly observed HST ACS field in IC~1613 (rectangle, upper
right).
The optical center of the galaxy is indicated by the white cross.
The two dashed ellipses correspond to the core radius (r$_c$) at 4\farcm5 ($\sim$ 1.0 kpc)
and the half-light radius (r$_h$) at 6\farcm5 ($\sim$ 1.4 kpc).  As can be seen from the
figure, the HST ACS field is located between the two.
Also indicated are the positions of the two previous HST WFPC2 fields (chevrons) from 
\citet{cole1999} (inner) and \citet{skillman2003a} (outer).
\label{f1}}
\end{figure}

\section{OBSERVATIONS AND DATA REDUCTION}\label{secred}

The ACS observations of IC~1613 were obtained between August 18 and 20, 2006. 
The F475W and F814W bands were selected as the most efficient combination to trace
age differences at old ages, since they provide the
smallest relative error in age and metallicity in the main-sequence and sub-giant 
regions (see C.\ Gallart et al. 2013, in prep).
Total integration times were 31,489 s in F475W and
27,119 s in F814W.  The observations were organized into 12 visits of 2 orbits
each, and each orbit was split into one F475W and one F814W exposure (in 
order to maximize sampling of variable star light curves). The observing
log is reported in \citet{bernard2010}. 
Dithers of a few pixels between exposures were introduced to
minimize the impact of pixel-to-pixel sensitivity variations (``hot pixels'') 
in the CCDs. The observed field of IC~1613 is
shown in Figure \ref{f1}.  At the adopted distance to IC~1613, the footprint of
the ACS covers 0.56 kpc$^2$. The optical scale length of IC~1613 is 
2\farcm9 \citep{bernard2007} and the stellar distribution can be traced out
beyond 15 arcminutes, so the 3\farcm4 $\times$ 3\farcm4 format of the ACS covers 
only a fraction of the area of IC~1613 ($\sim$ 9\%).

We analyzed the images taken directly from the STScI pipeline (bias, flat-field,
and image distortion corrected). Two PSF-fitting photometry packages,
DAOPHOT/ALLFRAME \citep{ste1994} and DOLPHOT \citep{dol2000},  
were used independently to obtain the photometry of the resolved stars.
Non-stellar objects and stars with discrepant and large uncertainties were 
rejected based on estimations of profile sharpness and goodness of fit. 
See \citet{mon_etal2010b} for more details about the photometry reduction
procedures.  
Individual photometry catalogs were calibrated using the equations provided 
by \citet{sir_etal2005}.  
The zero-point differences between the two sets of photometry are small 
($\lesssim$ 0.04 mag) and typical for obtaining HST photometry with different 
methods \citep{hill98, holtzman06}. 
We direct the reader to extensive photometry reduction comparisons of LCID
observations discussed in \citet{mon_etal2010b} and \citet{hidalgo2011}.  
For simplicity, the rest of this paper is based on only the DOLPHOT photometry 
dataset which contains 165,572 stars.

Signal-to-noise limitations, detector defects, and stellar crowding can all
impact the quality of the photometry of resolved stars with the resulting
loss of stars, changes in measured stellar colors and magnitudes, and
systematic uncertainties. To characterize these observational effects, we injected
$\sim 5\times 10^5$ artificial stars in the observed
images and obtained their photometry in an identical manner as for the real stars.
\citet{mon_etal2010b} and \citet{hidalgo2011} provide detailed descriptions of the procedures
we adopt for the characterization and simulation of these observational effects.

\begin{figure}[h]
\includegraphics*[width=8.6cm]{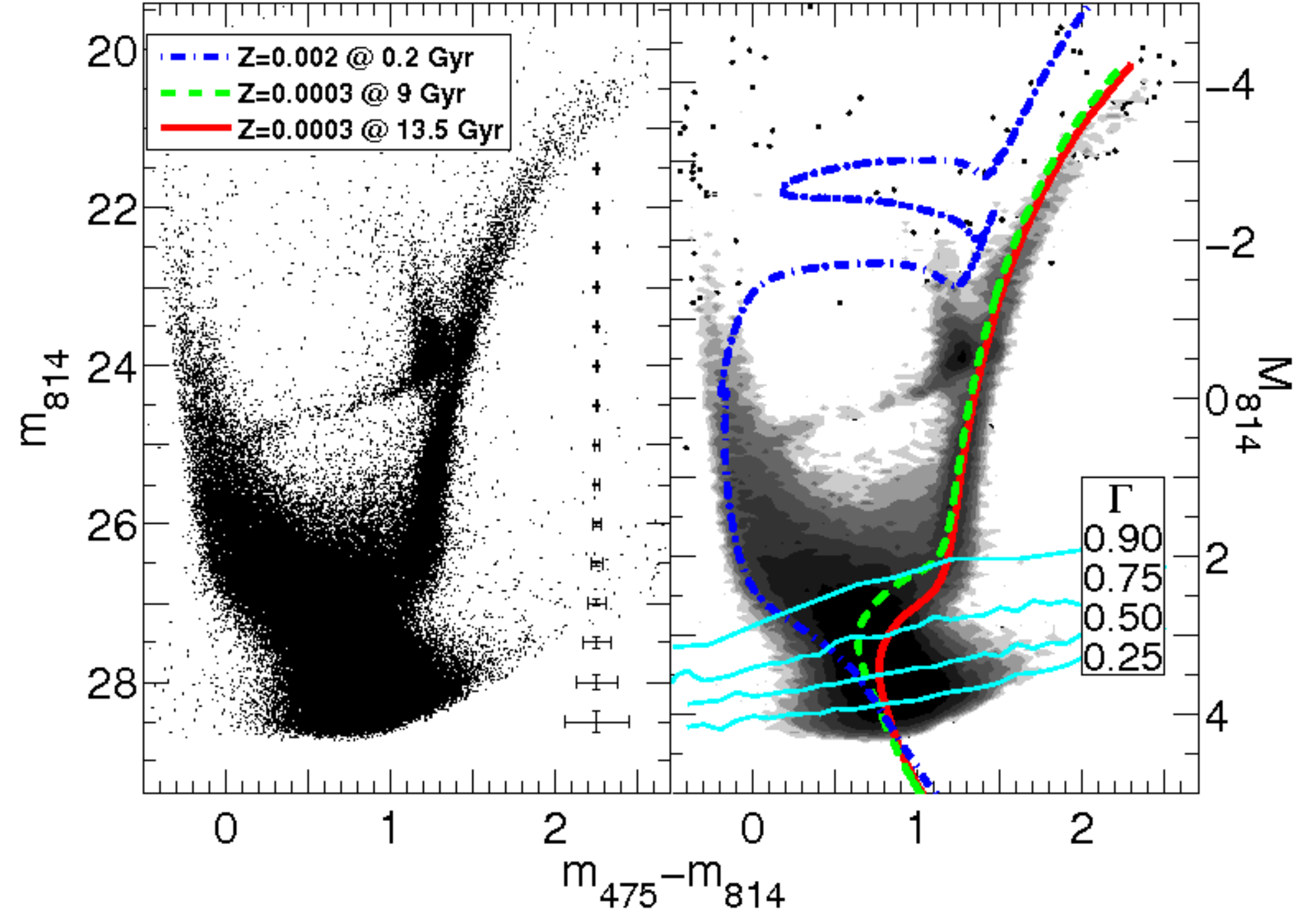}
\caption[ ]{The HST/ACS color-magnitude diagram of IC~1613 based
on 165,572 stars. In the left panel,
individual stars are plotted while the right panel shows star density levels.
Star densities bordering the plotted levels run from 8 to 512 stars dmag$^{-2}$,
evenly spaced by factors of 2. The left axis shows observed magnitudes corrected for extinction.
The right axis shows absolute magnitudes. A distance modulus of $(m - M)_0=24.40$ and
an extinction $A_{F475W}=0.094$, $A_{F814W}=0.047$ have been used. The lines across the
bottom of the right panel show the 0.25, 0.50, 0.75 and 0.90 completeness levels.
Three isochrones from the BaSTI stellar evolution library have been over plotted
for comparison.
\label{f2}}
\end{figure}

The CMD of IC~1613 is shown in Figure \ref{f2}. 
Individual stars are plotted in the left panel and density levels 
are shown in the right panel. 
The left axis shows magnitudes in the ACS photometric system corrected for extinction. 
Absolute magnitudes are given on the right axis using the adopted values for 
the distance modulus ($(m-M)_0=24.40$) and extinctions ($A_{F475W} = 0.094$ 
and $A_{F814W} = 0.047$ mag) from \citet{bernard2010}.
The completeness factor, $\Gamma$, the rate of 
recovered artificial stars as a function of magnitude and color, is shown
in Figure \ref{f2}, right panel, at $\Gamma=0.25, 0.50, 0.75$ and 0.90.  
Finally, in order to highlight the main features of the CMD, 
three isochrones from the BaSTI stellar evolution library \citep{pie_etal2004} 
are also shown in the right panel as labeled in the inset.

As shown in Figure \ref{f2}, our photometry reaches below the 
oldest main-sequence turn-off at the 50\% completeness limit,
allowing for very strong constraints on the oldest epochs of star formation.
These observations are $\sim 1.5$ mag fainter 
than the deepest CMD previously obtained for this galaxy \citep{skillman2003a}.
Since there are no early isolated episodes of elevated SFR in the SFH, 
there is no obvious indication of the time resolution in the early SFH of 
IC~1613 shown in Figure \ref{f2}.  However, extensive modeling with similar LCID datasets 
has shown that observations of this quality yield a time resolution of $\sim$ 1 Gyr 
\citep{mon_etal2010b, mon_etal2010c, hidalgo2011}.  

Consistent with earlier observations, 
a comparison of the observations with the over-plotted 
isochrones and the presence of an extended horizontal branch indicate that a 
very old, very low metallicity stellar population is present in the galaxy. 
The gap produced in the 
HB by the RR-Lyrae variables and a red clump (RC) at the red end of the HB
are clearly defined. 
An MS with stars younger than 200 Myr is clearly apparent. Some blue and red core 
helium-burning stars might also be present above the RC 
($m_{814}< 22.4$; $0.5<(m_{475}-m_{814})<1.2$). The RGB bump can be also observed 
at $(m_{475}-m_{814})\sim 1.4$ and $m_{814}\sim 23.1$ mag;  
this broad feature can be explained by an extended SFH 
\citep{mon_etal2010a}.

\section{THE SFH OF IC~1613}\label{secsfhresult}

\subsection{A Representative Field}\label{repfield}

As can be seen from Figure \ref{f1}, the HST ACS field of view covers only a small
fraction of IC~1613 ($\sim$ 9\%).  The position of the field was chosen mainly as
a balance between optimizing the number of stars but minimizing the effects of crowding
at the photometric limit.  However, the resulting position lies just inside of
the half-light radius of IC~1613, which is quite fortunate.

Ideally, we would like to characterize the global SFH of IC~1613 from 
the observed field.  
Stellar population gradients are universal in dwarf galaxies and have 
been observed over a wide 
range of physical characteristics \citep{dohm1997, gallagher1998, tolstoy1998, dohm1998,
battinelli&demers2000, aparicioetal2000, aparicio&tikhonov2000,
hidalgoetal2003, magrinietal2003, leisyetal2005, battinelli&demers2006, 
demersetal2006, hidalgoetal2008, gallartetal2008, noeletal2009,
rys2011, grocholski2012}. 
In all cases, the gradients are in the sense that the mean age of the 
stellar population is younger toward the center of the galaxy
\citep[see discussion in][for further details]{hidalgo2013}.

Since the
observed field is close to the half-light radius, we make a type of 
``mean value'' assumption and assume that the SFH derived from
the observed field is a relatively good representation of the global SFH
for IC~1613.  
The balance between outer fields (with relatively more older stars) and 
inner fields (with relatively more younger stars) should not be too different from
the SFH of a field at the half-light radius.
For the rest of the paper the discussion will be based on 
this assumption and the SFH history of the observed field and the global SFH of
IC~1613 will be used interchangeably.  
This assumption is supported by the comparison of HST observations of IC~1613 
from different fields presented in \S\ref{comp3}.
Note that our main result, the absence of
a dominant star formation episode at early times, would only strengthen 
by the addition of more younger stars at smaller radii.  The opposite 
possibility (a relative surplus of older
stars in the inner parts of the galaxy) is constrained by the direct observation
of the number of RGB stars in the central field from the earlier HST/WFPC2 
observations \citep[][see discussion in section \S\ref{comp3}]{cole1999}. 

\subsection{Main Features of the SFH of IC~1613}\label{secsfhfeatures}

Following the other LCID studies, we first use the IAC method, consisting of
IAC-star \citep{apa_gal2004}, IAC-pop \citep{apa_hid2009}, and MinnIAC 
\citep{apa_hid2009, hidalgo2011} in order to solve for the SFH and AMR.
Details of the methodology can be found in \citet{hidalgo2011}. 
Specifically, the solution has been obtained using the BaSTI 
\citep{pie_etal2004} stellar evolutionary libraries and the bolometric 
corrections of \citet{bed_etal2005}, and the uncertainties are the 
statistical uncertainties and do not included certain systematics (e.g.,
the use of a different stellar library).

\begin{figure}[h]
\includegraphics[width=8.6cm,angle=0]{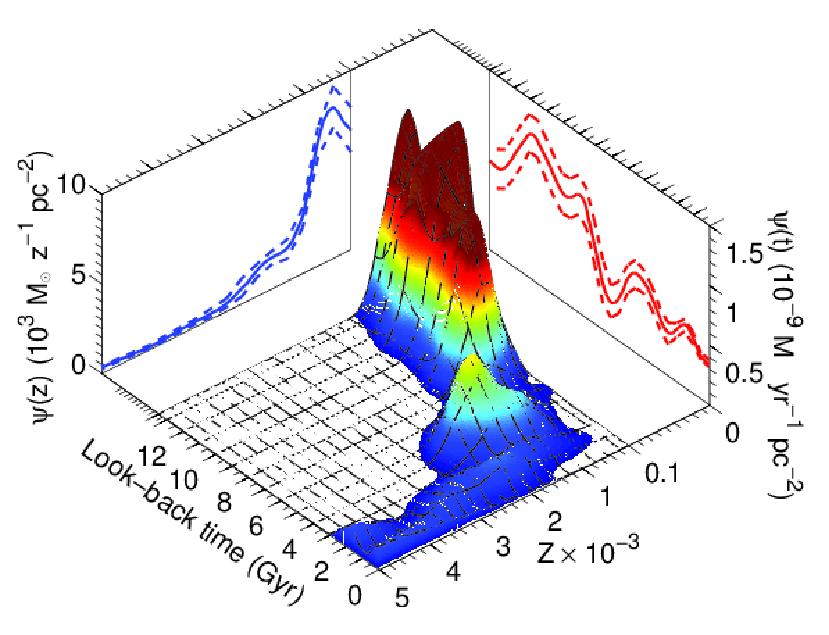}
\caption[ ]{Star formation history, $\psi(Z,t)$, of IC~1613 derived as described in
the text. The SFH as a
function of age $\psi(t)$ and metallicity $\psi(Z)$ are shown projected on the
$\psi-time$ and $\psi-metallicity$ planes, respectively. Dashed lines give the 
statistical error
intervals. The age-metallicity relationship is the projection onto the look-back
time--metallicity plane.
\label{f3}}
\end{figure}

Figure \ref{f3} shows a plot of the SFH of IC~1613, $\psi(t,Z)$, as a function of
both $t$ and $Z$.  $\psi(t)$ and $\psi(Z)$ are also shown
in the $\psi-t$ and $\psi-Z$ planes, respectively. The projection of $\psi(t,Z)$ on
the age-metallicity plane shows the AMR,  including the metallicity dispersion.

Figure \ref{f3} shows the presence of star formation at all times, and the gradual
build up of metallicity with time.  Note especially the relatively constant SFR over 
the first 6 Gyr of the history of IC~1613 (variations of $\sim$ 30\% or less in SFR).  
While the IAC-pop solution suggests a slight (factor of 2) decrease in 
average SFR about 8 Gyr ago ($z$ $\approx$ 1), there is no evidence for a dominant 
episode of star formation before the end of reionization
\citep[$\sim$ 12.8 Gyr, or $z$ $\approx$ 6,][]{fan2006}
or around the age of the peak SFR density of the universe 
\citep[$\sim$ 10 Gyr, or $z$ $\approx$ 2,][]{hopkins2004}).  
We emphasize that there is no evidence of a dominant
episode of star formation in very early times (lookback time $\ge$ 11 Gyr), 
despite our ability to resolve such a feature.

\begin{figure}[h]
\includegraphics[width=8.6cm,angle=0]{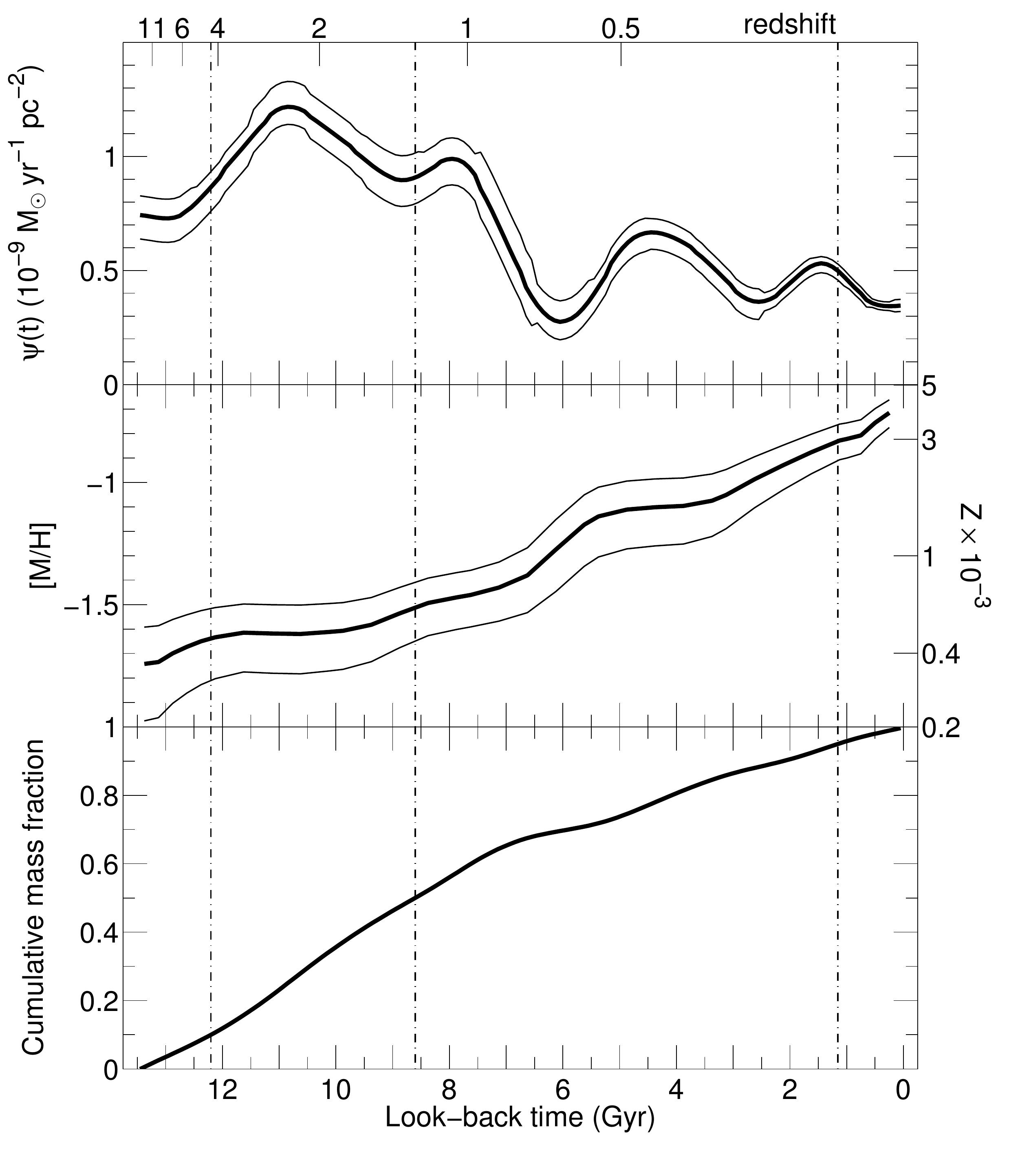}
\caption[ ]{Star formation rate, $\psi(t)$ (upper panel), metallicity (middle panel), 
and cumulative stellar
mass fraction (lower panel) of IC~1613. Thin lines give the uncertainties.
Vertical dotted-dashed lines indicate the times corresponding to the 10$^{\rm th}$,
50$^{\rm th}$ and 95$^{\rm th}$ percentiles of $\psi(t)$, i.e., the times for which
the cumulative fraction of mass converted into stars was 0.1, 0.5 and 0.95 of the
current total. The cumulative total mass of stars formed is $5.33\pm 0.10$ $10^6$ M$_\odot$.
The mean metallicity, $<$[M/H]$>$, is $-1.3\pm 0.1$.  The average star formation rate, 
$<\psi(t)>$, is $0.081\pm 0.001$ M$_\odot$ yr$^{-1}$.  A redshift scale is given in the upper axis. 
\label{f4}}
\end{figure}

In Figure \ref{f4}
we show $\psi(t)$, the AMR (with their associated errors), and the cumulative 
stellar mass fraction of IC~1613, as a function of time. Three vertical dashed lines show 
the ages of the 10th, 50th, and 95th-percentile of the integral of $\psi(t)$. 

The main features of the SFH of IC~1613 are nearly continuous star formation and
an accompanying smooth increase in the stellar metallicity.
The SFR varies by at most a factor of two from its mean value over the entire
lifetime of the galaxy.
Most importantly, there is no strong initial epoch of star formation, as is 
seen in the LCID dSph galaxies and transition (dSph/dIrr) galaxies.
This is reflected in the cumulative stellar mass fraction plot which is close to
the diagonal line of a constant SFR.

\begin{figure}[h]
\includegraphics[width=8.6cm,angle=0]{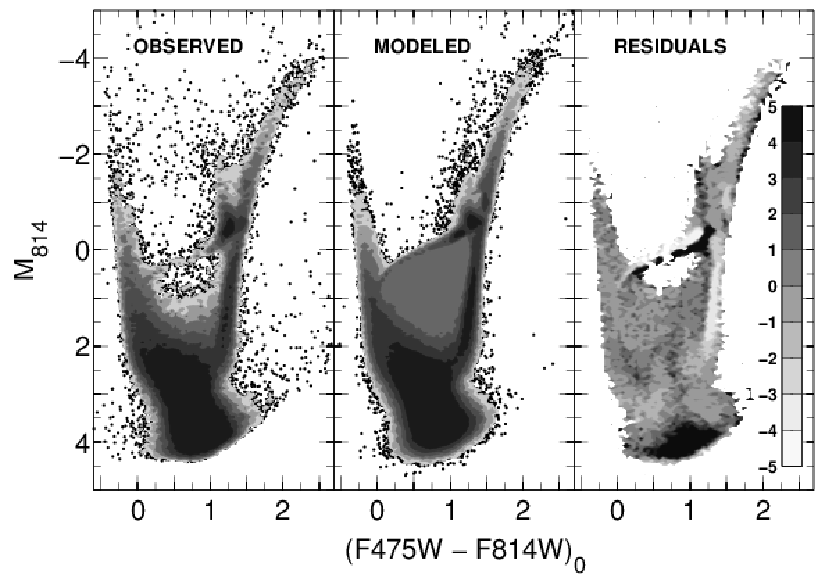}
\caption[ ]{Observed (left panel), calculated model (central panel), and
residual (right panel) CMDs. The calculated CMD has been built using IAC-star with
the solution SFH of IC~1613 as input. The grey scale and dot criteria for the first
two is the same as for Figure \ref{f2}. The residuals are in units of Poisson
uncertainties.\label{f5}}
\end{figure}

Figure \ref{f5} shows the observed CMD, the best-fit CMD, and the corresponding residuals.
Residuals are given in units of Poisson uncertainties, obtained as
$(n^o_i-n^c_i)/\sqrt{n^c_i}$, where $n^o_i$ and $n^c_i$ are the number of stars in bin
$i$ of an uniform grid defined on the observed and calculated CMDs, respectively.
Because the solution is heavily determined by the main sequence stars 
\citep[i.e., the RGB and HB stars are not used in the solution, 
see discussion in][]{hidalgo2013}, the residuals along the MS are negligible.
There are significant residuals at the faintest levels (where the incompleteness is large) 
and in the horizontal branch (which is typically not well fit by models).

\subsection{Comparison of SFHs Derived with Different Methods}\label{comresult}

\begin{figure}[h]
\includegraphics[width=8.6cm,angle=0]{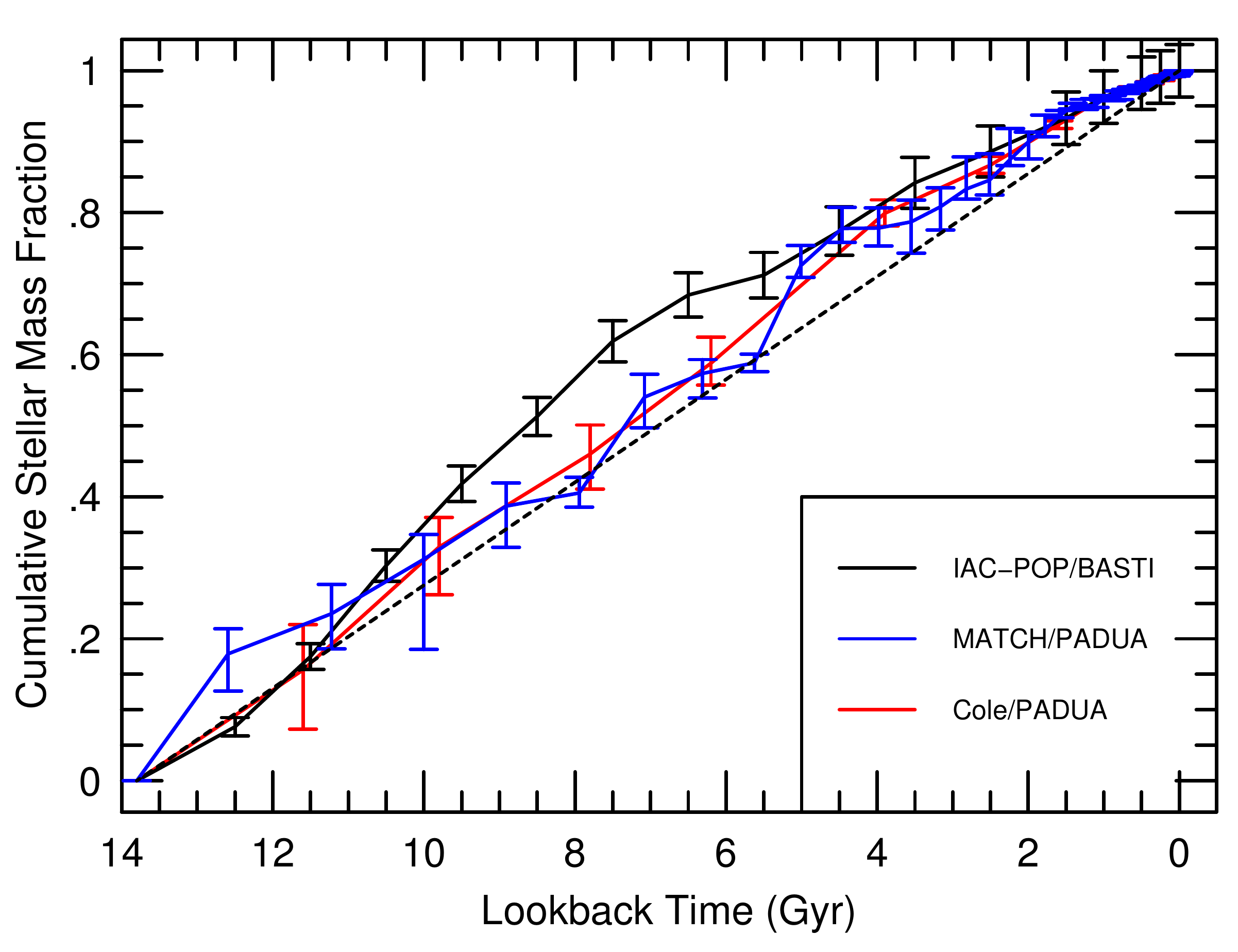}
\caption[ ]{Cumulative stellar mass fraction as a function of look-back time of the
solutions obtained for IC~1613 with different methods, indicated in the label.
IAC-Pop corresponds to the solution obtained with
IAC-star/MinnIAC/IAC-pop using the BaSTI stellar evolution library. 
Cole and Match methods make use of the \citet{girardi2010}
stellar evolution library.
\label{f6}}
\end{figure}

As for the other galaxies in the LCID sample, we have obtained the SFH of IC~1613 using 
the MATCH \citep{dol2002} and Cole \citep{skillman2003a} methods in addition 
to the IAC method. These two methods use \citet{girardi2010} for the stellar 
evolution library
for this comparison \citep[for a description of the main features of 
these methods and the particulars of their application to LCID 
galaxies, see][]{mon_etal2010b}.
In this exercise, we have allowed all parameters (e.g., distance, reddening, etc.) to 
be solved for optimally independently.  That is, we are not trying to compare
codes, but rather, trying to make an assessment of the systematic errors 
that may arise due to choices of different stellar libraries, codes, etc.

There are three main sources of uncertainties 
in deriving SFHs \citep[see, e.g., discussion in][]{apa_hid2009}: 
the input observational uncertainties, the statistical
uncertainties of the solution, and the systematic uncertainty of the limitations
of our knowledge of stellar evolution.
In a specific test of creating synthetic photometry from one stellar library and
using a different stellar library to derive a SFH, \citet{apa_hid2009} showed that
the uncertainties associated with our knowledge of stellar evolution dominated
those of the numerical methodology or the observation uncertainties.
They recommend always using more than one stellar evolution library when analyzing 
a real population to test for these effects.
\citet{weisz2011} and \citet{dolphin2012} have also shown that, for sufficiently 
deep photometry, the stellar evolution uncertainties, as approximated by 
differences between stellar evolution model and their resulting libraries, 
represent the dominant systematic uncertainty for derived SFHs. 
\citep[See also the discussion in][concerning the effects of varying heavy element 
abundance patterns.]{dotter2007}

Thus, using different stellar
libraries is integral to this comparison.  The SFH derived using MATCH
features uncertainties calculated following the prescriptions in \citet{dolphin2013}.
In this case, random uncertainties were generated using a hybrid Monte Carlo (HMC) process 
\citep{duane1987}, with implementation details as described by \citet{dolphin2013}.  
The result of this Markov Chain Monte Carlo routine is a sample of 10,000 SFHs
whose density is proportional to the probability density.  (That is, the density of
samples is highest near the maximum likelihood point.)  Upper and lower random error bars
for any given value (e.g., cumulative stellar mass fraction at a particular point in time) are
calculated by identifying the boundaries of the highest-density region containing 68\% 
of the samples, with the value 68\% adopted as it is the percentage of a normal 
distribution falling between the $\pm$ 1 $\sigma$ bounds.

Figure \ref{f6} shows the cumulative stellar mass 
fraction of IC~1613 as obtained with the three different methods. 
The agreement between the three methods is very good and all three
are best described as consistent with nearly constant star formation.
The one small difference between the SFHs is the offset at intermediate ages 
between the solution based on the BASTI models and the two based
on the Padua models.  A comparison using the same code with two
different stellar libraries shows that about half of this difference
is due to the differences between the stellar libraries (which 
have differences in physical inputs and differ in the  
treatment of core convective overshoot during the central H burning stage) 
and the other half is likely attributable to how different sections of the
CMD are binned and weighted.  

This comparison strongly supports the conclusion that the star 
formation in IC~1613 has been relatively constant for the lifetime of the galaxy.
The similar behavior of the three solutions allows us to be confident that 
our conclusions are independent of the stellar evolution libraries and
SFH solution method. Most importantly,
an early dominant star formation event is clearly ruled out.  

Note also that IC~1613 has clearly formed the vast majority of its stars after the 
epoch of reionization which occurred in the first Gyr of the history of
the universe \citep[cf.][]{fan2006}.

\subsection{Comparison of SFHs Derived from Other HST Observations of IC~1613}
\label{comp3}

\begin{figure}[h]
\includegraphics[width=8.6cm,angle=0]{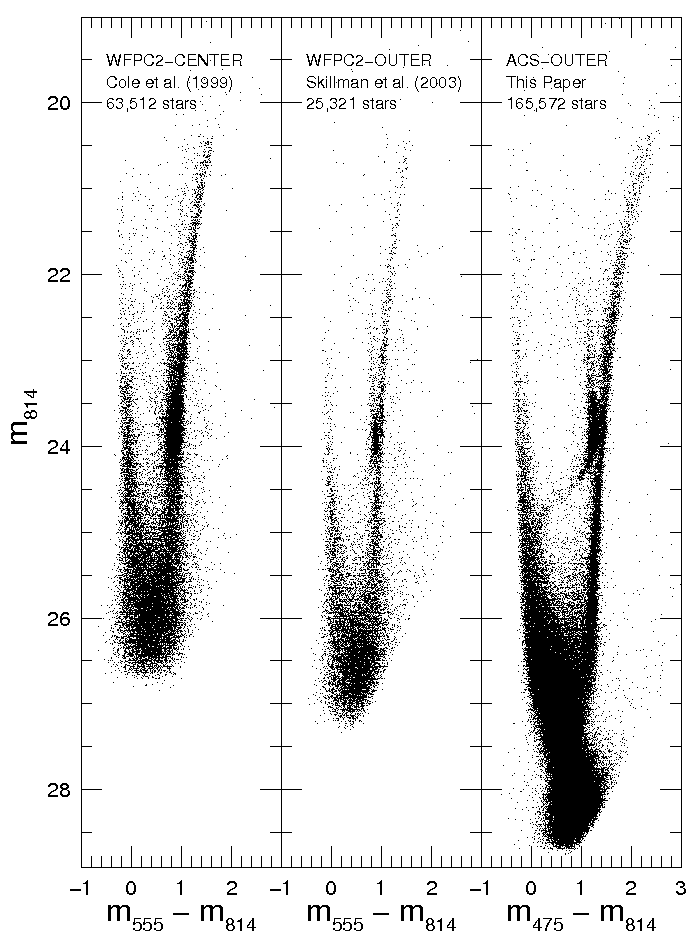}
\caption[ ]{Comparison of the photometry 
obtained for IC~1613 from three different fields observed with the HST.
WFPC2-CENTER corresponds to the 9 orbit observation presented in
\citet{cole1999},
WFPC2-OUTER corresponds to the 24 orbit observation presented in
\citet{skillman2003a}, and
ACS-OUTER corresponds to the new ACS observations presented here.
\label{f7}}
\end{figure}

\begin{figure}[h]
\includegraphics[width=8.6cm,angle=0]{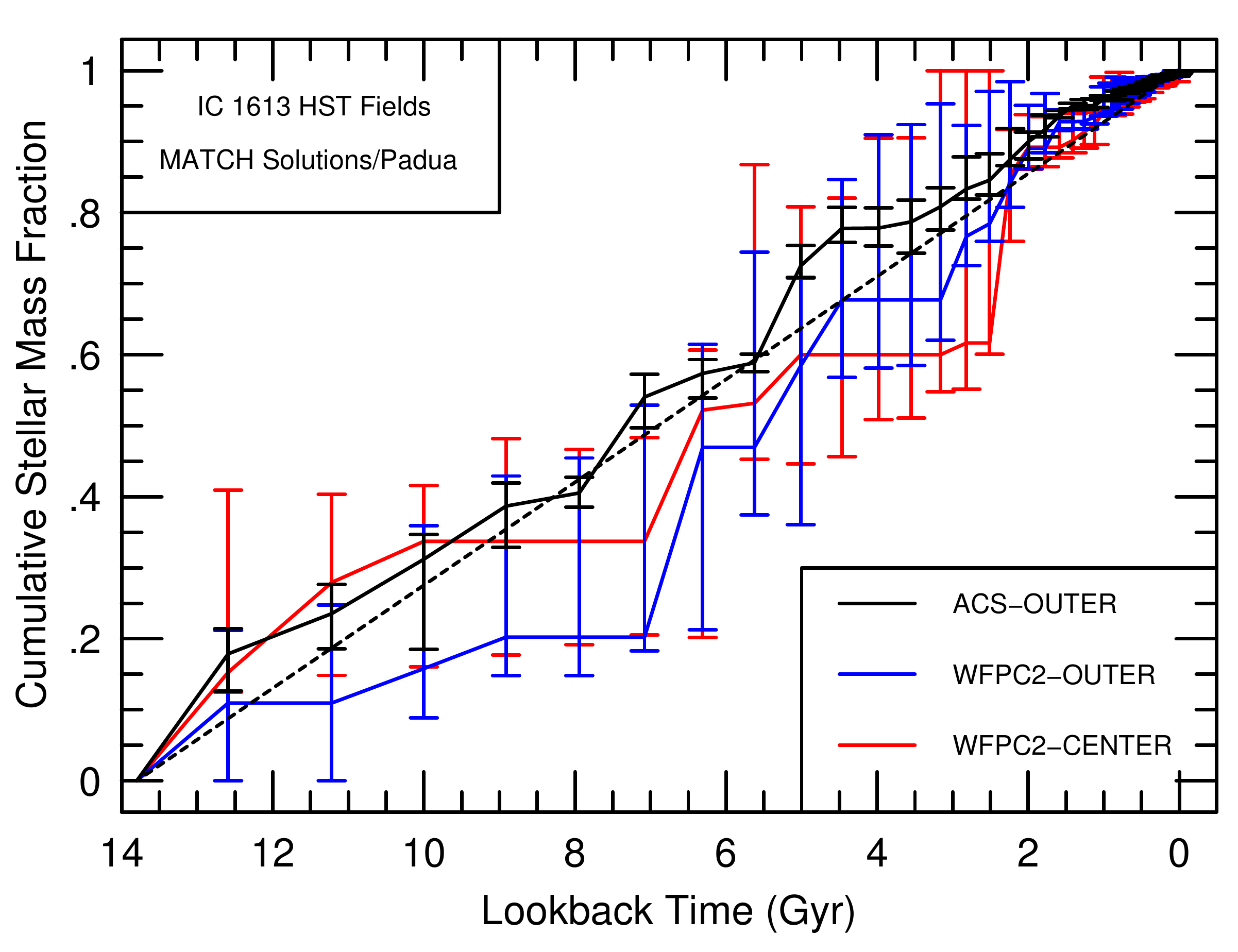}
\caption[ ]{Cumulative stellar mass fraction as a function of look-back time of the
solutions obtained for IC~1613 from three different fields observed with the HST.
All three solutions were derived with MATCH using the \citet{girardi2010} 
stellar evolution library. The errorbars were derived following the 
prescription outlined in \citet{dolphin2013}.
As in Figure \ref{f7},
WFPC2-CENTER corresponds to the 9 orbit observation presented in 
\citet{cole1999},
WFPC2-OUTER corresponds to the 24 orbit observation presented in 
\citet{skillman2003a}, and
ACS-OUTER corresponds to new ACS observations presented here.
Note that the deeper ACS observations have allowed significantly smaller
uncertainties at the earliest times.
All three SFHs are consistent with a constant SFR over
the lifetime of the galaxy, but only the deeper ACS observations clearly
rule out a dominant episode of early star formation.
\label{f8}}
\end{figure}

Relatively deep imaging of IC~1613 has been obtained with the HST three times:
a central field obtained with the WFPC2 camera (9 orbits) with a depth in I-band
of $\sim$ 26.5 mag.\ presented by \citet{cole1999},
a deep (24 orbits) outer field (radius $\approx$ 1.4 kpc) obtained with the 
WFPC2 camera with a depth in I-band of $\sim$ 27.3 presented 
by \citet{skillman2003a}, and our new imaging of an outer field
(radius $\approx$ 1.2 kpc) obtained with the ACS camera with a depth in I-band
of $\sim$ 28.7.  Figure \ref{f7} presents a comparison of the photometry for these 
three observations.  In making Figure \ref{f7}, identical quality cuts were
made on the data, so that they are directly comparable (although the WFPC 
observations were obtained with an F555W filter and the present ACS observations
were obtained with an F475W filter).

Figure \ref{f8} shows the cumulative stellar mass
fraction of IC~1613 as obtained from the three different observations.
The SFHs for the WFPC2 fields are newly derived from the archival data
\citep{dolphin2005, weisz2014}.
All three SFHs were obtained using the MATCH code \citep{dol2002} and the Padua
\citep{girardi2010} stellar evolution library.
The errorbars were derived following the
prescription outlined in \citet{dolphin2013}.

This comparison to SFHs derived from shallower data needs to be approached
carefully.  Clearly, the larger uncertainties inherent in the shallower observations
are reflected by larger errorbars on the individual points.  However, since 
the dominant component of the uncertainties are systematic, and since these can
only be estimated and not calculated directly, these should be interpreted as
``best estimates''.  Predicting early SFHs from photometry which does not reach the 
oldest main sequence turn-offs is vulnerable to systematic biases.  There is no 
substitute for the deep observations with
the resulting strong constraints on the old age SFH.  Nonetheless, this
comparison allows us to test our assumption of the representative nature of the 
ACS observation for the global SFH of IC~1613.

The main impression from Figure \ref{f8} is that the SFHs for all three observations
are best described as consistent with nearly constant star formation,
but only the deeper ACS observations clearly
rule out a dominant episode of early star formation.
Thus, Figure \ref{f8} does provide strong support for the assumption
that the SFH derived from the ACS observations can be assumed to be
representative of the global SFH for IC~1613.
Given the caveats from the previous paragraph we will refrain from any
further discussion of this comparison.

\section{The SFH of IC~1613 Compared to Other LCID Galaxies}\label{seclcid}

In this section, we will compare the SFH of IC~1613 with that of Leo~A
\citep[the other dIrr galaxy of the LCID sample,][]{cole2007},
Phoenix and \objectname[]{LGS-3}  
\citep[the two dSph/dIrr transition galaxies,][]{hid_etal2009, hidalgo2011}
and with those 
of Cetus and Tucana \citep[the two dSphs,][]{mon_etal2010b,
mon_etal2010c}.
Note that the LCID sample covers a range in dynamical masses.
From \citet{kirby2014} the dynamical masses of IC~1613, Leo~A,
LGS-3, Cetus, and Tucana are:
1.1 $\pm$ 0.2 $\times$ 10$^8$ $M_{\odot}$,
1.5 $^{+0.6}_{-0.5}$ $\times$ 10$^7$ $M_{\odot}$,
2.7 $^{+3.7}_{-2.0}$ $\times$ 10$^7$ $M_{\odot}$,
4.0 $^{+1.0}_{-0.9}$ $\times$ 10$^7$ $M_{\odot}$, and
7.1 $\pm$ 1.2 $\times$ 10$^7$ $M_{\odot}$, respectively.
For Phoenix, we derive a value of $\sim$2.7 $\times$ 10$^7$ $M_{\odot}$,
assuming a velocity dispersion of $\sim$8 km s$^{-1}$ \citep{zaggia2011}
and a half-light radius of 454 pc \citep{mcc2012}.

\begin{figure}[h]
\includegraphics[width=8.6cm,angle=0]{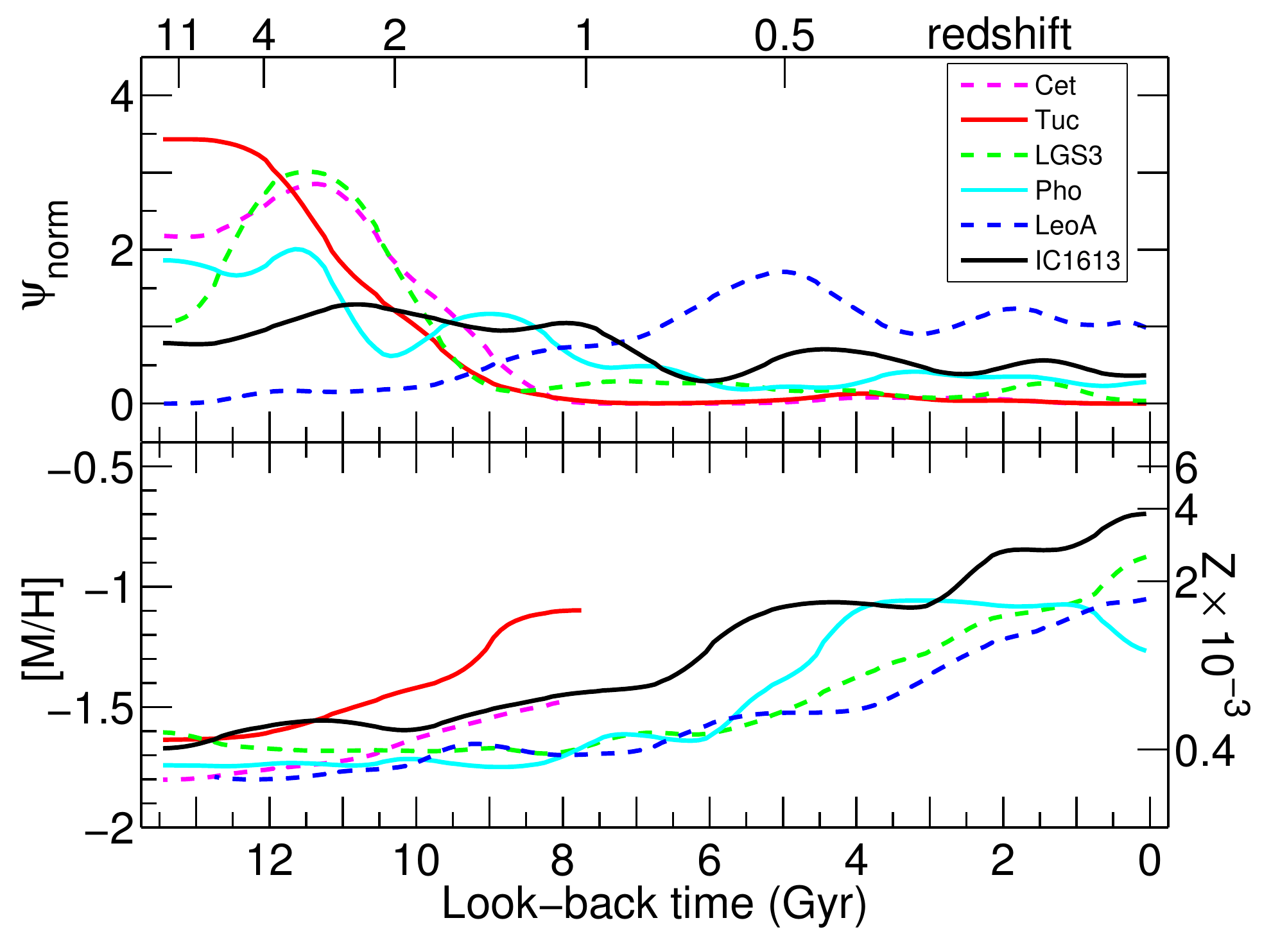}
\caption[ ]{Comparison between the SFHs and the AMRs of IC~1613
(this paper) and the other LCID galaxies. A redshift scale is given
on the top axis. 
\label{f9}}
\end{figure}

Figure \ref{f9} compares the SFH and the AMR of IC~1613 with the other LCID galaxies.
In the top panel of Figure \ref{f9}, the SFRs ($\Psi_{norm}$) have been normalized 
such that the area under each curve is 1 (i.e., the integral of the SFR over the
lifetime of the galaxy is 1). 
The differences between the SFHs as a function of morphological type is 
quite remarkable.
Both dIrr galaxies lack the dominant initial episode of star formation seen in 
both the dSph and transition type galaxies.
Although the SFHs are the main topic of this study, the AMRs are also 
solved for in deriving the SFHs and are included here for completeness.
Both dIrr galaxies also show relatively little metal enrichment during the 
first $\sim$ 6 Gyr, with gradual enrichment following.  

Note that \citet{hidalgo2011} highlighted the lack of early chemical
enrichment in the transition galaxy LGS-3 and hypothesized that the 
differences between the transition galaxies and dSphs (which show 
early chemical enrichment) might be understood if the dSphs 
were initially more massive systems than the transition galaxies.  
Under this hypothesis, 
the delayed metal enrichment in the transition galaxies is
due to higher losses of metals during times of higher SFRs for the 
lower mass transition galaxies.  
In this light, it is interesting that IC~1613 (and also Leo~A)
show delayed metal enrichment, but with no initial high SFR.
In this regard, further comparisons of the SFH and AMRs of 
transition galaxies and dIrr galaxies would be be great interest\footnote{
It is important to note that the definition and nature of transition galaxies is
not a completely settled issue.  Clearly, many transition galaxies are low mass
dIrrs lacking \ion{H}{2} regions simply due to a gap in recent massive star formation
as noted by \citet{skillman2003b}.  A nearby example of a galaxy with all
of the properties of a dIrr that has been labeled a transition galaxy is Pegasus
\citep{skillman1997}.
\citet{weisz2011} hypothesize that the {\it majority} of transition galaxies are simply
lower mass dIrrs \citep[supported by the recent study of Leo~T,][]{weisz2012}, 
but with a sub-sample of very gas poor galaxies like Phoenix \citep{young2007}.
If Phoenix and \objectname[]{LGS-3} are not representative of the typical
transition galaxy, then this point needs to be explored further before generalizations 
can be made about the true nature of the transition galaxies.}.

\begin{figure}[h]
\includegraphics[width=8.6cm,angle=0]{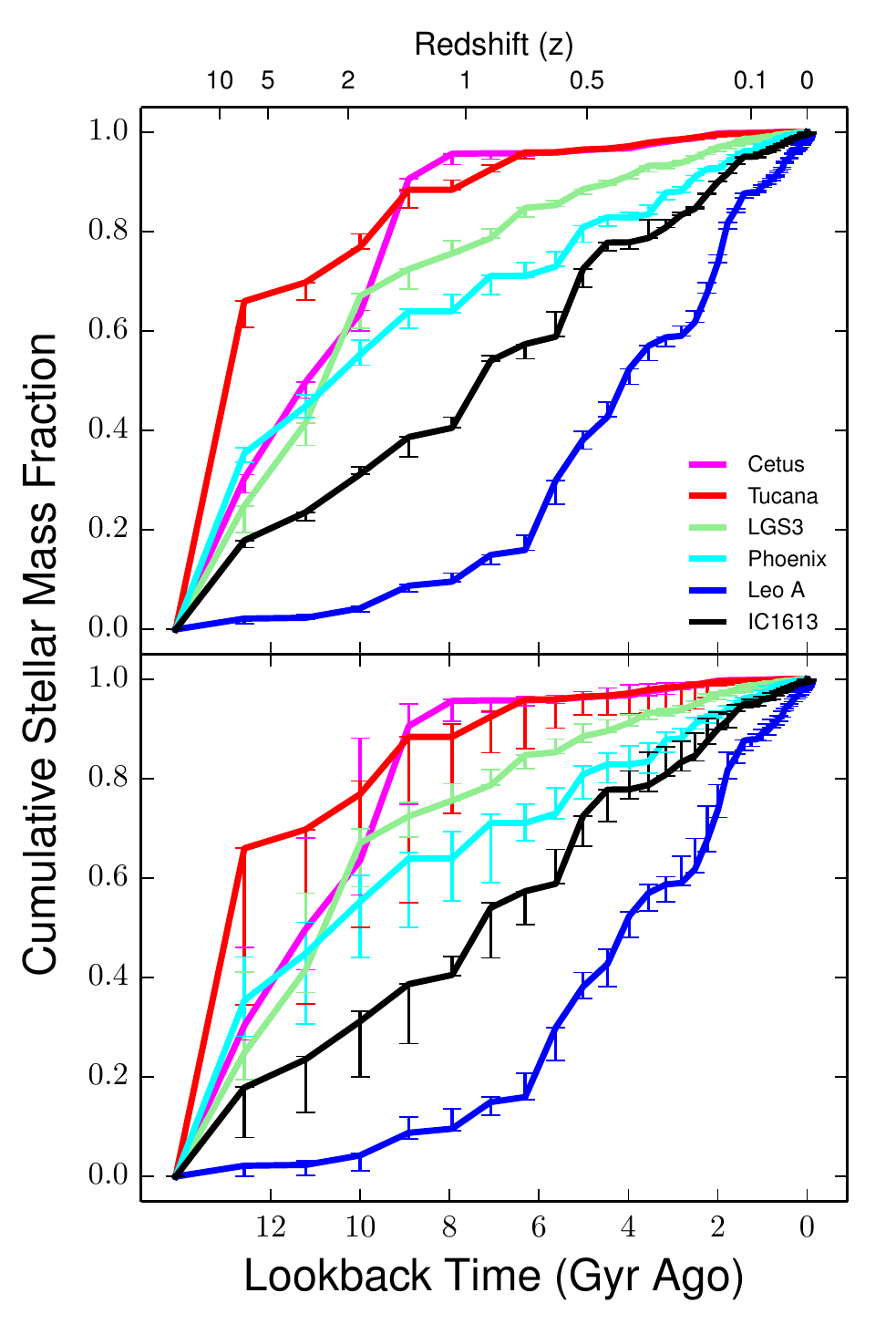}
\caption[ ]{Comparison between the SFHs of the LCID galaxies shown as cumulative
stellar mass fraction. The upper panel shows only the statistical uncertainties,
The lower panel accounts for estimated systematic uncertainties as discussed in
\citep{dolphin2012, dolphin2013}. A redshift scale is given on the top axis.
\label{f10}}
\end{figure}

Before comparing to theoretical models, we introduce one last comparative figure.
In Figure \ref{f10}, we show a comparison between the SFHs of the six LCID galaxies as 
cumulative stellar mass fractions.  In the upper panel, only the statistical
uncertainties are shown.  In the lower panel, the systematic uncertainties are
included following the methodology of \citet{dolphin2013}.  Because all of the LCID galaxies
have been observed to comparable depth, systematics in the models should have similar
impacts to all of the galaxies.  Thus, it is likely appropriate to make comparisons
using the upper panel.  Nonetheless, the lower panel shows the larger uncertainties
encountered when trying to account for systematics and, even with the larger uncertainties,
the six galaxies are shown to each have distinctive features in their SFHs.

Portraying the SFHs as cumulative stellar mass fractions (as opposed to 
SFR as a function of time) is the optimal way to compare
observations to theoretical models for several reasons.  Variations
in observed SFRs can be strongly affected by time binning and the changing time
resolution as a function of lookback time.  Often, it is possible to have very 
different impressions of a single SFH simply by changing the time binning.
It is possible to match the observational time binning by reducing the resolution 
in the models, but using the cumulative stellar mass fraction as the diagnostic
avoids this problem altogether.  It is also possible to compare galaxies at any 
arbitrary value of the cumulative stellar mass fraction, as opposed to choosing
particular values to focus on.  In the comparisons that follow, we will use the
cumulative stellar mass fraction as the sole diagnostic.  Note that there is one
obvious failing of the cumulative stellar mass fraction as the sole diagnostic, and 
that is the lack of information about the absolute masses of the systems. In the 
following comparisons, we will provide information about the masses of both
the observed and modeled systems.

\section{The EARLY EVOLUTION OF IC~1613 AND THE OVER-COOLING PROBLEM}\label{seccosmo}

\subsection{Background}

Recently, \citet{orban08}, \citet{sawala2011}, \citet{weinmann2012}, and \citet{kuhlen12} 
have all highlighted the difficulty of producing dwarf galaxies in simulations 
with properties comparable to those observed in the current universe.  
Together, the introductions to their papers give a comprehensive overview of the 
problems with modeling dwarf galaxy evolution.  

To summarize, there are two major problems.  The first problem is the observed abundance
of low-mass galaxies.  The observed slope of the low-mass galaxy luminosity function is shallow
relative to the slope of the halo mass function and this difference seems to be a result of 
an extreme inefficiency of galaxy-scale star formation over cosmic times. 
The well known problem of
``the missing satellites'' \citep{kwg93, kly_etal1999, moo_etal1999}
is one manifestation of this problem at very low luminosities/halo masses.

The second problem relates to {\it when} stars are formed in galaxies. A natural 
assumption is that the time-scale for global star formation is related to the 
time-scale of baryonic accretion onto galaxies. However, low-mass halos assemble 
almost all of their (dark matter) mass at high redshift \citep[e.g.,][]{fakhouri2010}, 
while essentially all field dwarf galaxies show star formation continuing to the 
present day \citep[e.g.,][]{weisz2011}. IC~1613 and Leo~A are extreme examples of this, 
with essentially a constant SF rate across cosmic time for IC~1613 and delayed star
formation in the case of Leo~A.

In order to suppress the abundance of low-mass galaxies, most theoretical models 
impose strong feedback in small halos \citep[e.g.,][]{mac_fer1999, gnedin2000, 
bul_etal2001, sto_etal2002, kra_etal2004, ric_gne2005, str_etal2008, 
saw_etal2010, busha2010, sawala2013}.
Two processes can dramatically affect the formation and evolution 
of dwarf-sized halos: heating from the ultraviolet radiation arising from 
cosmic reionization and feedback from internal supernovae. Both processes are, 
in principle, capable of completely halting the star formation in a dwarf halo and 
even fully removing all of the galaxy's gas. 
Employing these feedback mechanisms while tying star formation to the collection
of baryons has the effect of predicting that essentially all star formation in low-mass 
halos happens at early times. 
Both semi-analytic models and hydrodynamical
models fail to satisfactorily reproduce the evolution of low mass galaxies in
the sense that stellar mass is over-produced per dark matter halo mass.
Stars are produced too quickly at early times, 
resulting in stellar mass fractions that are too high
by an order of magnitude.  Model galaxies usually do not have the 
high gas mass fractions commonly observed in present day dwarfs \citep[e.g.,][]{begum2008},
and, as a result, star formation falls off too fast 
and the colors of simulated dwarf galaxies are too red at $z$ $=$ 0.
This is a manifestation of the ``over-cooling''
problem which is a challenge for all modeling efforts, but which 
is exaggerated at low masses.  

The missing satellites problem is largest at the lowest masses, and IC~1613 is
massive enough that it could be expected to emerge from reionization with
its gas intact.  Thus, IC~1613 does not provide as strong a test of models for
this problem as the other, less massive, galaxies in the LCID sample. 
However, our deep HST observations have produced SFHs for IC~1613 and Leo~A with small 
uncertainties even at earliest times and have completely ruled out the possibility
of an early, dominant phase of star formation.
{\it This represents a real challenge for any model where star formation 
follows mass assembly.}

\subsection{Direct Comparison with Models Highlighting the ``Over-Cooling'' Problem}

\begin{figure} 
\includegraphics[width=8.6cm,angle=0]{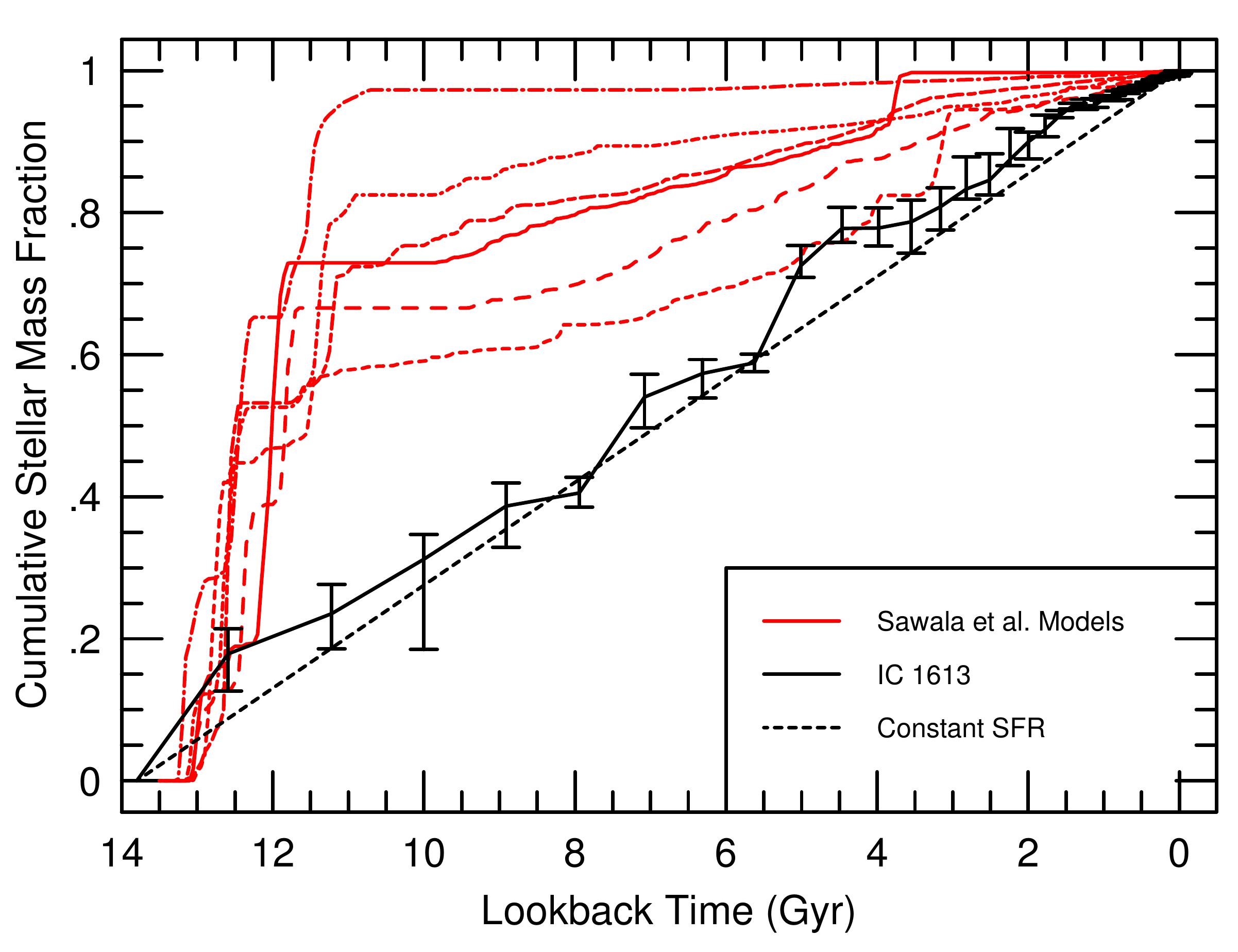}
\caption[ ]{Comparison of the SFH of IC~1613 with model SFHs calculated by
\citet{sawala2011}.  The six model galaxies have halo masses at $z$ $=$ 0 of
$\sim$ 10$^{10}$ M$_{\odot}$, which is comparable to the estimated
halo mass for IC~1613.
The difference in the early SFHs between the models and IC~1613 is very clear. 
}
\label{f11}
\end{figure}

\citet{sawala2011} have highlighted the shortcomings of models for dwarf galaxies
where star formation follows directly from mass assembly. 
They presented a set of high-resolution  
hydrodynamical simulations of the formation and evolution of isolated dwarf galaxies 
including the most relevant physical effects, namely metal-dependent cooling, 
star formation, feedback from Type II and Ia SNe, and UV background radiation. 
Their results are very useful for a direct comparison with observations. 
They identify and study six halos with present day dark matter halo masses 
of $\approx$ 10$^{10}$ M$_\odot$ 
with stellar masses ranging between
$4.9\times 10^7$ M$_\odot$ and $1.0\times 10^8$ M$_\odot$.
For comparison, IC~1613 has an estimated stellar mass of $\sim$
10$^8$ M$_\odot$ \citep{mcc2012}, and the observed maximum rotation velocity
of 25 km s$^{-1}$ \citep{lake1989}. If we associate this peak rotation
velocity with the maximum circular velocity of IC~1613's host dark matter
halo, this corresponds to a virial mass\footnote{Using data from the
Millennium-II Simulation \citep{boylan2009}, we
find that: \\
$V_{\rm max}=36\,{\rm km\,s^{-1}}\,(M_{\rm vir} / 10^{10}\,M_{\odot})^{0.316}$.} 
of $M_{\rm vir}=3.2 \times 10^{9}\,M_{\odot}$. 
This is likely to be a lower limit, however, as the peak of the circular velocity 
curve for the dark matter halo may be attained at
radii beyond the extent of the stellar or gaseous tracers. Abundance matching
models find that galaxies with stellar masses of $10^{8}\,M_{\odot}$ should be
hosted by halos with virial masses of $\sim 3\times 10^{10}\,M_{\odot}$ 
\citep[e.g.,][]{behroozi2013}.
Given these estimates, the models of Sawala et al. provide an excellent sample
to compare to the derived SFH of IC~1613.

As Figure \ref{f11} demonstrates clearly, the early evolution of the model galaxies 
all depart significantly from that of IC~1613.  In essence, the models
all have SFRs which depend directly on the gas content of the galaxies.
As highlighted by \citet{sawala2011}, since, in the current paradigm, most of the 
mass of a low mass galaxy is in place well before $z$ $=$ 1 \citep[e.g.,][]{fakhouri2010}, 
then any prescription in which star formation follows gas content is going to 
build up most of the stellar mass before $z$ $=$ 1.

This discrepancy between models and observations, while demonstrated here for 
IC~1613, is beginning to appear to be the norm. 
Although the numbers are very limited, so far, all gas-rich dwarfs for which
there exist sufficiently deep HST observations 
\citep[i.e., Leo A, Leo T, SMC:][]{cole2007, weisz2012, weisz2013}
show no evidence for a dominant early episode of star formation.

\subsection{Potential Solutions for the ``Over-Cooling'' Problem}

There have been various attempts to improve models to better accommodate the
inefficient star formation in dwarf galaxies.
\citet{stinson2007} used SPH + N-body simulations and showed that supernova 
feedback could disrupt enough gas to temporarily quench star formation. 
Episodic star formation follows from the cycling of gas into a hot halo
with subsequent cooling and infall. 
\citet{orban08} account for the prolonged star formation observed in
dwarfs by adding a stochastic variation in the density threshold of the 
star formation law.  Essentially, this simply reduces the efficiency of
star formation by hand, but the result is a significantly improved match to
the observed SFHs of dwarfs.
Recently, \citet{stinson2013} have demonstrated that the thermal feedback from
early star formation can effectively decouple star formation from mass
assembly, thus producing more realistic SFHs.

\citet{gnedin2009} and \citet{gnedin2010, gnedin2011} have proposed a different physical
approach by investigating the conversion of atomic to molecular gas and its
affect on the efficiency of star formation. 
This is well motivated by the observational work of \citet{leroy2008}
and \citet{bigiel2008}, which clearly demonstrate that star formation
follows the molecular gas content and not the total gas content. 
In the specific case of dwarf galaxies, \citet{kuhlen12} show that models 
which incorporate a star formation 
prescription regulated by the local abundance of molecular hydrogen lead to 
the less efficient star formation that is desired.
Unfortunately, their first attempts are not completely successful as
they state, ``like most cosmological galaxy formation simulations to date, 
our simulated galaxies suffer from the so-called baryonic overcooling problem, 
resulting in unrealistically high central densities (and hence strongly peaked 
circular velocity curves) and stellar mass fractions in our high-mass halos 
that are too large compared to observations.''  However, this  
appears to be a very promising avenue for future exploration 
\citep[see also][]{christensen2012, zolotov2012, kuhlen2013}.

\citet{stark2013} used a semi-analytic galaxy formation model to investigate 
the properties of the satellites of Milky Way-like galaxies and 
were able to match the star formation histories of several dwarf satellites.
The extended SFHs of these satellites were a result
of a gas density threshold for star formation.  Galaxies could have large reservoirs
of gas lying just below the threshold, leading to inefficient star formation
(requiring a minor accretion or interaction event).

\begin{figure} 
\includegraphics[width=8.6cm,angle=0]{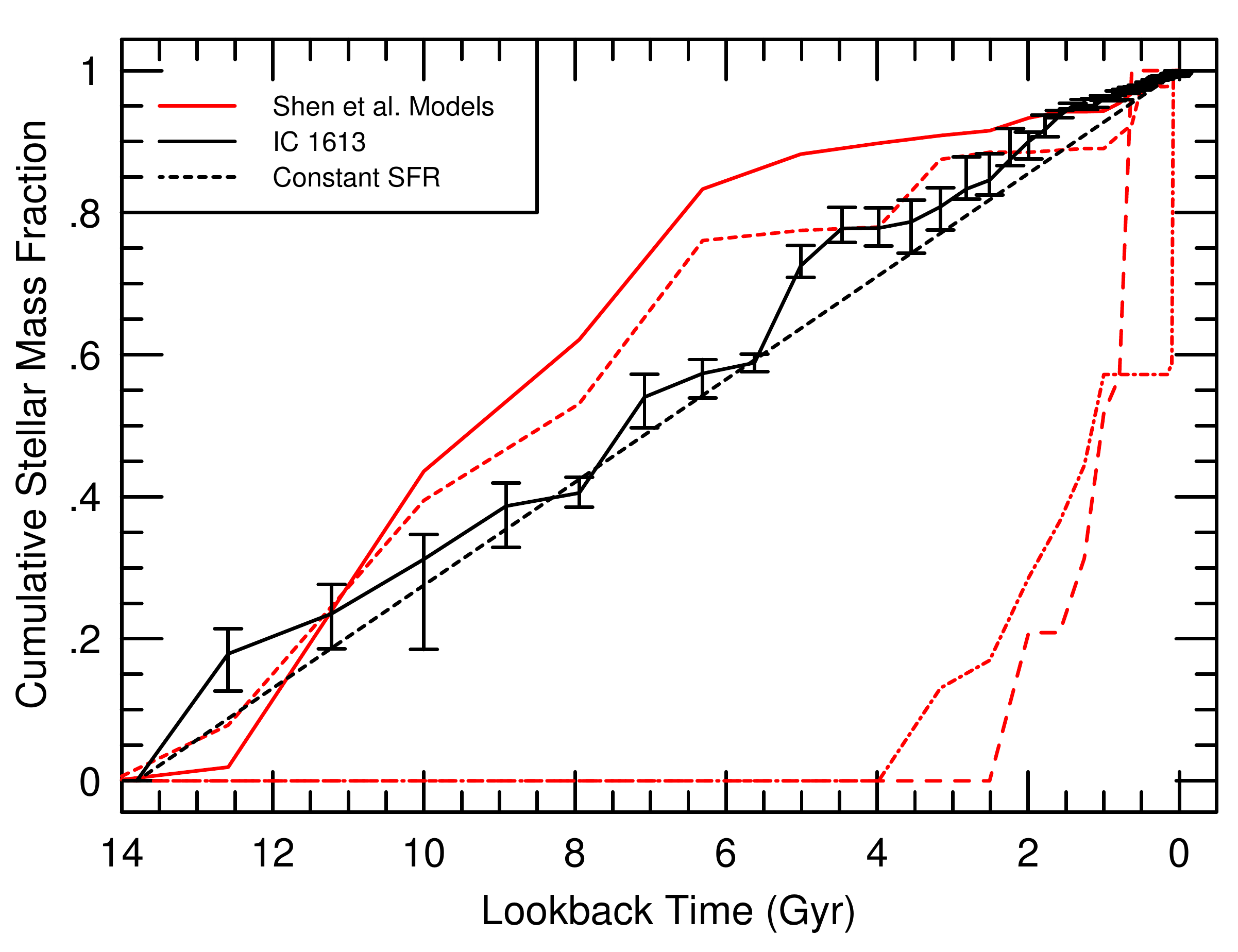}
\caption[ ]{Comparison of the SFH of IC~1613 with model SFHs calculated by
\citet{shen2013}.  The two model galaxies with nearly constant star formation
are ``Bashful'' and ``Doc'' with halo masses at $z$ $=$ 0 of
$\sim$ 4 $\times$ 10$^{10}$ M$_{\odot}$ and $\sim$ 1 $\times$ 10$^{10}$ M$_{\odot}$,
respectively, comparable to the estimated
halo mass for IC~1613.
``Bashful'' and ``Doc'' have nearly constant star formation and lack a dominant
early episode of star formation as is observed in IC ~1613.
The two model galaxies with a delayed onset of star formation
are ``Dopey'' and ``Grumpy'' with smaller halo masses at $z$ $=$ 0 of
$\sim$ 3 $\times$ 10$^{9}$ M$_{\odot}$ and $\sim$ 2 $\times$ 10$^{9}$ M$_{\odot}$,
respectively, (which are considerably less massive than IC~1613, but might be more
directly comparable to Leo~A).
}
\label{f12}
\end{figure}

\citet{shen2013} have produced new simulations including supernova feedback, 
a star formation recipe based on a high gas density threshold, 
metal-dependent radiative cooling, 
turbulent diffusion of metals and thermal energy, and UV background radiation.
They reproduce the observed stellar mass and cold gas content, the 
SFHs, and the metallicities typical of field dwarfs in the Local Volume.
In Figure \ref{f12}, we reproduce the cumulative stellar mass fraction
for four of the model galaxies in \citet{shen2013}.
Two of the galaxies (``Bashful'' and ``Doc'') have virial masses which are
comparable to IC~1613.  These two models show nearly constant star formation 
with no dominant early episode of star formation, as observed in IC ~1613.  
In detail, the SFHs of 
these two models start later and rise faster during intermediate ages.  
The slower start of the models as compared to that of the measured SFHs may be
attributed to differences in how time is defined in the two methods.
That is, the simulations start at a fixed, given time, whereas
the observationally derived ages come from stellar evolution libraries, and
therefore depend on their underlying physics (e.g., nuclear reaction rates).

The two lower mass galaxies (``Dopey'' and ``Grumpy'') show no early star 
formation at all, and only start forming stars at intermediate ages.
These two models correspond to halo masses significantly less than that of IC~1613, 
but might be more comparable to that of Leo~A (cf. Figure \ref{f10}).
Note, however, that the stellar masses are even more discrepant.
That is, Dopey and Grumpy have stellar masses of 10$^5$ and 
5 $\times$ 10$^5$ M$_{\odot}$ compared to
10$^8$ and 6 $\times$ 10$^{6}$ M$_{\odot}$ for IC~1613 and Leo~A. 

Naively, the late onset of star formation in the Dopey and Grumpy models
would appear to be in contradiction to observations in which
all dwarf galaxies observed to date show evidence for an old  ($\ge$ 10
Gyr) population, particularly as evidenced by the presence of RR Lyrae 
stars.  For example, RR Lyrae stars have been observed in Leo~A 
\citep{dolphinea2002, bernard2013}. 
However, \citet{shen2013} hypothesize that such systems may be
the very metal poor systems necessary for the production of extremely 
metal deficient galaxies such as I~Zw~18 \citep[see, e.g., discussion and 
references in][]{skillman2013}.

\clearpage

\subsection{Other Implications}

One of the successes of recent modeling of dwarfs has been the ability of
simulations to solve the cusp/core problem for dwarf galaxies 
\citep[e.g.,][]{governato2010, pontzen2012, governato2012, teyssier2013}.
The dark matter halos of observed dwarf galaxies show nearly constant density cores,
while steep central dark-matter profiles are expected from CDM models
\citep[e.g.,][]{deblok2001}.
The resolution to this conundrum lies in strong outflows from supernovae 
to remove low-angular-momentum gas from the centers of dwarf galaxies.
This solves two problems as bulgeless dwarf galaxies are formed with
shallow central dark-matter profiles, and a large fraction of the original
baryons are lost from the galaxy, resulting in the lower baryon fractions
(relative to the universal fraction) which are observed
\citep[e.g.,][]{mcgaugh2010}.

What is not clear is whether the current ideas concerning how to 
produce inefficient star formation also lead to cored dark matter halos.
Clearly the strong outflows need to occur when the mass assembly is 
taking place, which is early on.  If the vast majority of the star formation
is delayed until later times, after the mass assembly is virtually finished,
then the central outflow solution to the cusp/core problem may not arise
self-consistently.  A related
point is the energetic requirements for creating bulgeless galaxies and forming
cores in state-of-the-art hydrodynamic simulations. Simulations that are
currently successful require (1) highly efficient coupling between the energy
released in a supernova explosion and the ISM of a galaxy -- values typically
adopted are 0.4--1 \citep{governato2010, teyssier2013, shen2013} and 
(2) an artificial prevention of gas cooling for approximately
$10^7$ years in a region surrounding a supernova explosion, meant to capture
processes that may occur at sub-grid scales and cannot be directly included in
the simulation \citep{stinson2006}. The success of simulations in producing
bulgeless, cored dwarf galaxies with relatively low-level, extended star
formation can be therefore seen as predictions for the effective coupling of
supernova explosions with the ISM and for the efficacy of additional feedback
processes (such as stellar winds) at heating the ISM on small scales. These
predictions of very efficient coupling may be confirmed or refuted with future
generations of simulations that model the relevant processes directly.

Finally, although we have emphasized the 
difficulty of producing a nearly constant SFR as seen in 
IC~1613, observations of dwarf galaxies in the Local Volume point toward
a large diversity of SFHs. Determining the critical 
parameters which drive that diversity is paramount.  How can it be that
dwarfs with halos of similar mass, which presumably have similar accretion
histories and early formation times, have vastly different SFHs?
Is environment and interaction history the main driver of the diversity
\citep[see, e.g.,][]{sawala2012, benitez2013}? From an observational point of view, 
strong constraints will depend on accurate determinations of the early
SFHs of a larger number of galaxies within and beyond the Local Group from
both deep photometry as presented here and spectroscopic studies of 
individual stars \citep[e.g.,][]{kirby2011, deboer2012a, deboer2012b}.
Clearly the sample of one presented here, and the very small number of
comparably observed galaxies presents a great limitation to our 
knowledge of the evolution of dwarf galaxies.

\section{SUMMARY AND CONCLUSIONS}\label{seccon}

We have presented the SFH of the dIrr galaxy IC~1613, based on deep 
HST photometry obtained with the ACS. 

\begin{itemize}

\item The SFH with relatively small uncertainties has been obtained for 
the entire lifetime of the galaxy.
The solution shows that the SFH of IC~1613 is consistent with
a constant SFR over its entire lifetime.

\item Most or all the star formation was produced in IC~1613 after the
reionization epoch, assumed to occur $\sim 12.8$ Gyr ago.  There is no 
evidence of an early dominant episode of star formation in IC~1613.

\item A comparison of the derived SFH of IC~1613 with the models of 
\citet{sawala2011} reinforce their observation that models where 
star formation follows mass assembly form too many stars too early.  
This well known aspect of
the so called ``over-cooling problem'' appears to be universal, but
now, with the deep HST photometry of IC~1613, we see that the problem 
reaches back to the very earliest times in the evolution of galaxies.

\item  There are proposed solutions to the over-cooling problem for
dwarf galaxies.  The solutions discussed in this paper rely on 
an efficient coupling of supernova feedback to the ISM.
The predictions of very efficient coupling may be confirmed or refuted 
with future generations of simulations that model the relevant 
processes directly and through future observations of dwarf galaxy SFHs
with similar quality to the LCID studies.

\end{itemize}

\acknowledgments

We would like to thank Till Sawala for sharing the numerical results of 
his modeling, and Greg Stinson, Piero Madau, Alyson Brooks, and the 
referee for helpful comments.
Support for this work was provided by NASA through grant GO-10515
from the Space Telescope Science Institute, which is operated by
AURA, Inc., under NASA contract NAS5-26555.
Support for DRW is provided by NASA through Hubble Fellowship grant
HST-HF-51331.01 awarded by the Space Telescope Science Institute.
The computer network at IAC operated under the Condor software license has been used. 
Authors SH, AA, CG, and MM are funded by the IAC (grant P3/94) and, with SC, PS, and EB,
by the Science 
and Technology Ministry of the Kingdom of Spain (grant AYA2007-3E3507) and
Economy and Competitiveness Ministry of the
Kingdom of Spain (grant AYA2010-16717). 
This research has made use of NASA's Astrophysics Data System
Bibliographic Services and the NASA/IPAC Extragalactic Database
(NED), which is operated by the Jet Propulsion Laboratory, California
Institute of Technology, under contract with the National Aeronautics
and Space Administration.

\end{document}